\documentclass{article}

\usepackage{arxiv}

\usepackage{amsmath}
\usepackage[utf8]{inputenc} 
\usepackage[T1]{fontenc} 
\usepackage{booktabs} 
\usepackage{amsfonts}  
\usepackage{nicefrac} 
\usepackage{microtype}  
\usepackage{graphicx}
\usepackage{cite}
\usepackage{doi}
\usepackage{svg}
\usepackage{mathtools}

\newcommand{\vc}[1]{{\mathbf{ #1}}}

\newcommand{\CC}{{\mathcal{C}}}

\newcommand{\olQ}{\overline{Q}}
\newtheorem{thm}{Theorem}{}
{}
{}
{}
\newtheorem{proof}{Proof}{}
\newtheorem{defn}{Definition}{}
\newtheorem{lem}[thm]{Lemma}
\newtheorem{eg}[thm]{Example}

\usepackage{tikz}
\usetikzlibrary{shapes, arrows, shadows, fit, patterns}

\DeclarePairedDelimiter{\norm}{\lVert}{\rVert}
\DeclareMathOperator{\dom}{dom}
\newcommand{\tr}{{\rm Tr}\,}

\title{Extensions on `A Convex Scheme for the Secrecy Capacity of a MIMO Wiretap Channel with a Single Antenna Eavesdropper'}

\author{ \href{https://orcid.org/0000-0001-9956-2801}{\includegraphics[scale=0.06]{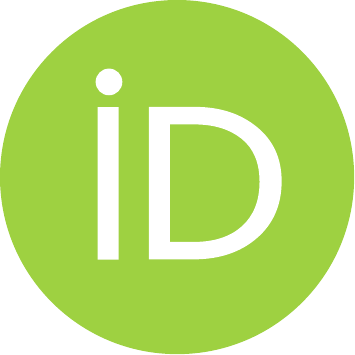}\hspace{1mm}Jennifer Chakravarty} \\
School of Mathematics\\
University of Bristol, United Kingdom\\
\texttt{jennifer.chakravarty@bristol.ac.uk} \\
\And
\href{https://orcid.org/0000-0002-3645-6670}{\includegraphics[scale=0.06]{orcid.pdf}\hspace{1mm}Oliver Johnson} \\
School of Mathematics\\
University of Bristol, United Kingdom\\
\texttt{oliver.johnson@bristol.ac.uk} \\
\AND
\href{https://orcid.org/0000-0002-4879-1206}{\includegraphics[scale=0.06]{orcid.pdf}\hspace{1mm}Robert Piechocki} \\
Department of Electrical \& Electronic Engineering \\
University of Bristol, United Kingdom \\
\texttt{robert.piechocki@bristol.ac.uk} 
}

\date{}

\renewcommand{\shorttitle}{Extensions on `A Convex Scheme for the Secrecy Capacity of a MIMO Wiretap Channel with a Single Antenna Eavesdropper'}

\hypersetup{
pdftitle={Extensions on `A Convex Scheme for the Secrecy Capacity of a MIMO Wiretap Channel with a Single Antenna Eavesdropper'},
pdfsubject={cs.IT},
pdfauthor={Jennifer Chakravarty, Oliver Johnson, Robert Piechocki},
pdfkeywords={NOMA, MIMO, Physical Layer Security},
}
\pgfdeclarelayer{bg}    
\pgfsetlayers{bg,main}  
\tikzstyle{decision} = [diamond, draw, fill=blue!20, text width=4.5em, text badly centered, node distance=3cm, inner sep=0pt]
\tikzstyle{block} = [rectangle, draw, fill=blue!20, text width=6em, text centered, rounded corners, minimum height=4em]
\tikzstyle{line} = [draw, -latex']
\tikzstyle{cloud} = [draw, ellipse,fill=red!20, node distance=3cm, minimum height=2em]
\tikzstyle{clear} = [rectangle,fill=white, node distance=3cm, minimum height=2em, text width=5em, text centered]

\begin{document}
\maketitle
\begin{abstract}

One key metric for physical layer security is the secrecy capacity. This is the maximum rate that a system can transmit with perfect secrecy. For a Multiple Input Multiple Output (MIMO) system (a newer technology for 5G, 6G and beyond) the secrecy capacity is not fully understood.
For a Gaussian MIMO channel, the secrecy capacity is a non-convex optimisation problem for which a general solution is not available.

Previous work by the authors showed that the secrecy capacity of a MIMO system with a single eavesdrop antenna is concave to a cut off point. In this work, which extends the previous paper, results are given for the region beyond this cut off point. It is shown that, for certain parameters, the presented scheme is concave to a point, and convex beyond it, and can therefore be solved efficiently using existing convex optimisation software.

\end{abstract}


\section{Introduction}\label{sec:intro}

 Multiple-input multiple-output (MIMO) systems play a large role in achieving higher capacities and thus are central in 5G technologies, with `massive' MIMO being a central technology for 5G and future wireless~\cite{hoydis2013massive}.
Security for any modern day system is vital however there are several fundamental questions which remain open with regards to the physical layer security of a MIMO channel when compared to the equivalent model for point to point single antenna systems. Indeed, the secrecy capacity for a Gaussian MIMO wiretap channel is one of these open problems. The work in this chapter aims to addresses this, contributing a theorem which gives a region where a MIMOSE channel has a concave secrecy capacity equation. Knowing when the equation is concave allows for the problem to be efficiently solved, giving the secrecy capacity and thus the maximum rate for secure communications for the given channel.

This paper extends the results of \cite{ICC_CJP}, where it was shown that there exists a concave region in the secrecy capacity function for the single eavesdrop antenna case. The theoretical set up and notation are all the same. In \cite{ICC_CJP}, an upper bound was found for the secrecy capacity. It was shown that up to this limit the formulation was convex, allowing the use of low complexity convex optimisation methods to find the, previously unknown, secrecy capacity. In this paper, we find a lower bound, extending the region of known convexity and the usefulness of these results.

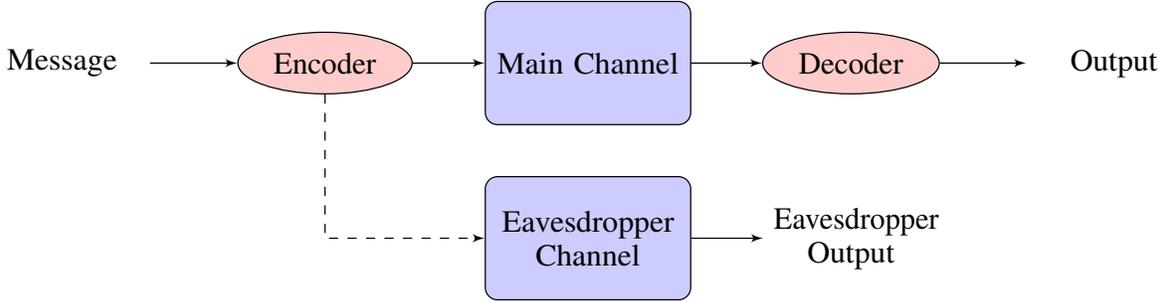
\begin{figure}[h]
    \begin{center}
    \resizebox{\textwidth} {!} {
        \begin{tikzpicture}
            \tikzset{node distance = 2cm}
            \node [block] (channelm) {Main Channel};
            \node [cloud, left of=channelm] (encoder) {Encoder};
            \node [cloud, right of=channelm] (decoder) {Decoder};
            \node [block, below of=channelm] (channele) {Eavesdropper Channel};
            \node [clear, left of=encoder] (source) {Message};
            \node [clear, right of=decoder] (outputm) {Output};
            \node [clear, right of=channele] (outpute) {Eavesdropper Output};
            \path [line] (source) -- (encoder);
            \path [line] (decoder) -- (outputm);
            \path [line] (channele) -- (outpute);
            \path [line] (encoder) -- (channelm);
            \path [line,dashed] (encoder) |- (channele);
            \path [line] (channelm) -- (decoder);
        \end{tikzpicture}
        }
    \end{center}
    \caption{The Wiretap Channel \cite{wyner}.}
    \label{fig:wiretap}
\end{figure}

Physical layer security has an information theoretic foundation and is theoretically unbreakable. Quantifying security in terms of information leakage was first considered by Shannon in \cite{shannon_secrecy} and the traditional model stems from Wyner's work in 1975 \cite{wyner}, the `Wiretap Channel' seen in Figure \ref{fig:wiretap}.
The typical set up considered involves two legitimate users, Alice and Bob, transmitting across a channel with an eavesdropper, Eve.
The information theoretic constructs give an idea of how much useful information the eavesdropper is able to obtain, known as the information leakage.
These secrecy measures, which depend on block length and the channel quality are independent of computational power and thus applicable to any technologies.

Multiple antenna systems play a large role in achieving higher capacities and thus are central in 5G technologies, with `massive' MIMO being a central technology for 5G and future wireless~\cite{hoydis2013massive}.
Security for any modern day system is vital however there are several fundamental questions which remain open with regards to the physical layer security of a MIMO channel when compared to the equivalent model for point to point single antenna systems.
Indeed, the secrecy capacity for a Gaussian MIMO wiretap channel, is one of these open problems.
The work in this chapter aims to addresses this, contributing a theorem which gives a region where a MIMOSE channel has a concave secrecy capacity equation.
Knowing when the equation is concave allows for the problem to be efficiently solved, giving the secrecy capacity and thus the maximum rate for secure communications for the given channel.

\subsection{Theoretical setup}
\begin{figure}
    \centering
    \begin{tikzpicture}[scale=1.3]
    \draw (0,0) rectangle (2,3.75);
    \node at (1,1.8) {Alice};
    \draw [thick] (2,0.5) -- (2.5,0.5) -- (2.5,1) -- (2.25,1.25);
    \draw [thick] (2.5,1) -- (2.75,1.25);
    \draw [thick, dotted] (2.375,1.5) -- (2.375,2.25);
    \draw [thick] (2,2.5) -- (2.5,2.5) -- (2.5,3) -- (2.25,3.25);
    \draw [thick] (2.5,3) -- (2.75,3.25);
    \draw (6,0.65+0.85) rectangle (8,4.40+0.85);
    \node at (7,2.525+0.85) {Bob};
    \draw [thick] (6,1.15+0.85) -- (5.5,1.15+0.85) -- (5.5,1.65+0.85) -- (5.25,1.9+0.85);
    \draw [thick] (5.5,1.65+0.85) -- (5.75,1.9+0.85);
    \draw [thick, dotted] (5.625,2.15+0.85) -- (5.625,2.9+0.85);
    \draw [thick] (6,3.15+0.85) -- (5.5,3.15+0.85) -- (5.5,3.65+0.85) -- (5.25,3.9+0.85);
    \draw [thick] (5.5,3.65+0.85) -- (5.75,3.9+0.85);    
    \draw (6,-2+0.85) rectangle (8,-0.25+0.85);
    \node at (7, -1.125+0.85) {Eve};
    \draw [thick] (6,-1.5+0.85) -- (5.5,-1.5+0.85) -- (5.5,-1+0.85) -- (5.25,-0.75+0.85);
    \draw [thick] (5.5,-1+0.85) -- (5.75,-0.75+0.85);
    \draw [thick, blue, dashed] (2.75,1.825) -- (5.25,2.525+0.85);
    \draw [thick, blue, dashed] (2.75,1.825) -- (5.25,-1.125+0.85);
    \end{tikzpicture}
    \caption{The MIMOSE wiretap channel}
    \label{fig:mimose}
\end{figure}
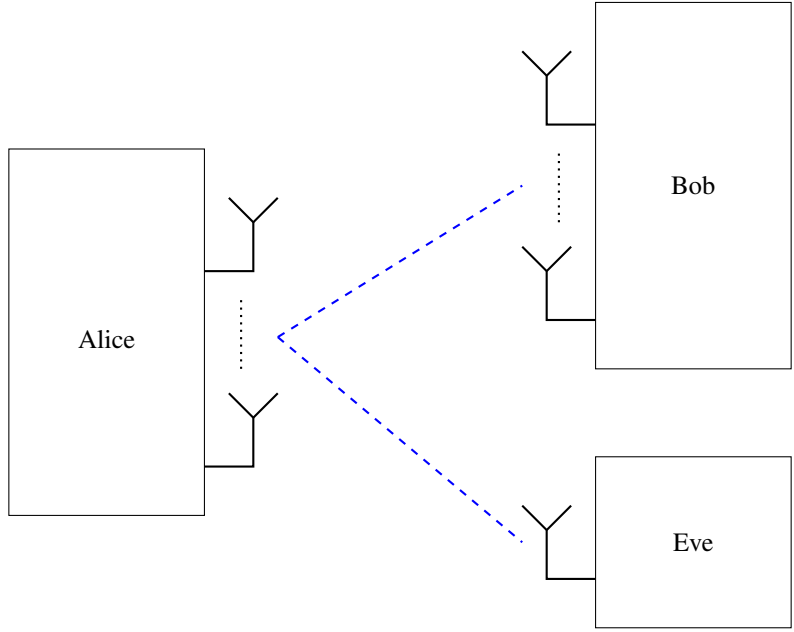

This work concerns a MIMO channel with $N_A$ transmit antennas and $N_B$ receive antennas at the legitimate receiver. The legitimate users, Alice and Bob, are communicating in the presence of a passive eavesdropper, Eve, with $N_E$ antennas. For the results of this chapter to hold, Eve has a single eavesdrop antenna, that is $N_E=1$ as depicted in Figure \ref{fig:mimose}.

The channel between the transmitter and the legitimate receiver shall be referred to as the \emph{main channel} while the channel between the transmitter and the eavesdropper shall be referred to as the \emph{eavesdropper channel}. Their channel matrices are described by $H_{B}$, an $N_{B} \times N_A$ matrix for the main channel and $H_{E}$, an $N_{E} \times N_A$ matrix for the eavesdropper channel.

The input signal, $\vc{x}$, is drawn from a distribution with zero mean and covariance matrix $Q\succeq 0$, which is a positive semidefinite matrix.
The received vectors at Bob and Eve, denoted $\vc{y}$ and $\vc{z}$ respectively, are:
\begin{align}
\vc{y}&=H_{B}\vc{x}+\vc{n}_{B}, \nonumber \\ 
\vc{z}&=H_{E}\vc{x}+\vc{n}_{E}. \nonumber
\end{align}
where $\vc{n}_{B}$ and $\vc{n}_{E}$ are the Gaussian noise vectors for the two channels
\begin{align*}
\vc{n}_{B}&\sim \mathbb{C}N(0,I_{N_{B}})
\end{align*}
and
\begin{align*}
\vc{n}_{E}&\sim \mathbb{C}N(0,I_{N_{E}})
\end{align*}
where each element of the noise vector is statistically independent. Similarly, the channel matrices are modelled with IID entries assuming independence between each antenna element.
The matrix $I_k$ denotes the identity matrix of size $k\times k$.
The input signal is subject to a power constraint $P$, meaning that the trace of the covariance matrix, $Q$, is bounded above by this quantity. That is,
$$\tr Q=\sum_{i=1}^{N_A}\mathbb{E}[\vc{x}_{i}\vc{x}_{i}^{*}]\leq P.$$
Without the power constraint above, the capacity is theoretically infinite, which does not provide much insight in a practical setting.

\subsection{Secrecy capacity}
The open problem we are addressing in this chapter is the secrecy capacity for the outlined system setup. The secrecy capacity, $C_{s}$, for the MIMO wiretap channel was established in~\cite{oggier_mimo_wiretap},~\cite{MISOME} and~\cite{MIMOME} to be of the form
\begin{align}\label{eq:ch_MIMO_seccap}
C_{s}= \max_{Q: \tr(Q)\leq P}\log\det(I_{N_{B}} + H_{B}QH_{B}^{*})-\log\det(I_{N_{E}} + H_{E}QH_{E}^{*})
\end{align}
where we note that, since the mean of the input signal is always zero, the maximum is being taken over all input distributions satisfying the power constraint.

The optimisation problem in Equation \eqref{eq:ch_MIMO_seccap} is not easily solved for $Q$ and the solution is only known for a subset of scenarios and remains open in the general case.
The difficulty lies in the fact that the optimisation is not convex and thus analytically challenging.
Knowing the optimal $Q$ is useful for a number of reasons, some of which are outlined below.

\begin{itemize}
    \item The mean of the input signal is always zero, so the covariance matrix, $Q$, is the characterising variable for the input distribution.
    \item The input covariance gives details for the optimal input scheme for secrecy and rate requirements.
    \item Knowledge of the optimum covariance matrix gives the true secrecy capacity.
    \item Once the secrecy capacity is known, any rate of transmission below this is secure by definition, giving a secure region for reliable rates of transmission.
\end{itemize}

The key contribution of this chapter is for the Gaussian MIMO wiretap channel with a \emph{single antenna eavesdropper}, a subset of the unknown MIMOSE family of wiretap channels. The secrecy capacity is examined for this open problem and a region is established where the problem is provably concave. The concavity of the problem gives an efficient method of determining the optimal input covariance matrix associated with the secrecy capacity of a system. The scheme given is valid for the MIMOSE channel where the receiver has at least as many antennas as the transmitter. That is, $N_B \geq N_A$ and $N_E=1$.

This family of antenna configurations overlaps with only two known cases, the point to point single antenna case where Alice, Bob and Eve each have one antenna, and the so called `(2,2,1)' case. Our results are compared with their results in Section \ref{sec:mimo_result_overlap}.

The theory of this chapter goes as follows: the secrecy capacity equation is reformulated into a problem which is convex, this allows existing convex optimisation tools and software to find the optimal solution to Equation \eqref{eq:ch_MIMO_seccap}.

The proof relies on properties of symmetric matrices and functions of the channel and thus for ease of notation we define the following positive semidefinite symmetric $N_A \times N_A$ matrices which are used in the statement of Theorem \ref{thm:main} and throughout the proof:
\begin{align} K_{B}&=(H_{B}^{*}H_{B})^{\frac{1}{2}}, \label{eq:kbdef} \\
K_{E}&=(H_{E}^{*}H_{E})^{\frac{1}{2}}. \label{eq:kedef}
\end{align}

\section{Concave region for the secrecy capacity}\label{sec:thm}
The key limitation in solving Equation \eqref{eq:ch_MIMO_seccap} is the fact that it is non-convex. In order to exploit existing convex solvers, we must first reformulate the secrecy capacity equation to an equivalent but tractable optimisation problem.
We know that $\log\det(\cdot)$ is known to be concave and twice differentiable for positive semidefinite arguments. It follows that each individual $\log\det(\cdot)$ term in Equation \eqref{eq:ch_MIMO_seccap} is concave.
This can be seen by considering their arguments. Since $Q$ is a covariance matrix, it is restricted to positive semidefinite matrices by definition. The identity matrix is trivially a positive semidefinite matrix and thus the terms
\begin{align*}I_{N_{B}} + H_{B}QH_{B}^{*}\end{align*}
and
\begin{align*}I_{N_{E}} + H_{E}QH_{E}^{*}\end{align*}
will also be positive semidefinite.
This means that both of the terms
\begin{align*}\log\det(I_{N_{B}} + H_{B}QH_{B}^{*})\end{align*}
and
\begin{align*}\log\det(I_{N_{E}} + H_{E}QH_{E}^{*})\end{align*}
are concave.
However, in general, their difference is neither convex nor concave. We will reformulate the problem in order to restrict the problem space to a region where the difference is concave. Broadly speaking, this is done by fixing the second term and then varying its value.
Hence we define the following problem:
\begin{align}\label{eq:opt_cvx}
    \max_{\tr(Q)\leq P} &\log\det(I_{N_{B}} + H_{B}QH_{B}^{*})-\log(s), \\
    \text{such that }&\quad s=\det(I_{N_{E}} + H_{E}QH_{E}^{*}) \nonumber \\
    \text{and }&\quad Q \succeq 0. \nonumber
\end{align}

The following work is constrained to a single eavesdrop antenna since, generally speaking, $\det(\cdot)$ is not a convex constraint.
When the problem space is limited in this way, the matrix argument $I_{N_{E}} + H_{E}QH_{E}^{*}$ is a scalar value.
Since $\log\det(I_{N_{B}} + H_{B}QH_{B}^{*})$ is concave and the maximisation is taken over a convex set, it can be seen that by fixing the value of $s$, this becomes a concave problem.

With $s$ fixed, Equation \eqref{eq:opt_cvx} is concave however it is no longer equivalent to Equation \eqref{eq:ch_MIMO_seccap}. In order to bridge this gap, we must vary our value of $s$ and take an overall maximum. This is the overarching idea which is formally laid out in the following section.

For the optimal value of $s$, Equation \eqref{eq:opt_cvx} is an equivalent problem to Equation \eqref{eq:ch_MIMO_seccap} and consequently will yield the same solution.

\subsection{Statement of theorem}
Each value of $s$ gives a separate convex optimisation problem in Equation \eqref{eq:opt_cvx}. For each optimisation, the output is a corresponding covariance matrix $Q$ and the maximum value of the argument. We aim to vary $s$ and take the maximum over each of the aforementioned outputs. 

We begin by defining functions of the input covariance matrix $Q$
\begin{equation}s(Q)=\det(I_{N_{E}}+H_{E}QH_{E}^{*})\end{equation} and \begin{equation}f(Q)=\log\det(I_{N_{B}}+H_{B}QH_{B}^{*})-\log s(Q).\end{equation}

\begin{figure}[h]
      \centering
      \includegraphics[width=\textwidth]{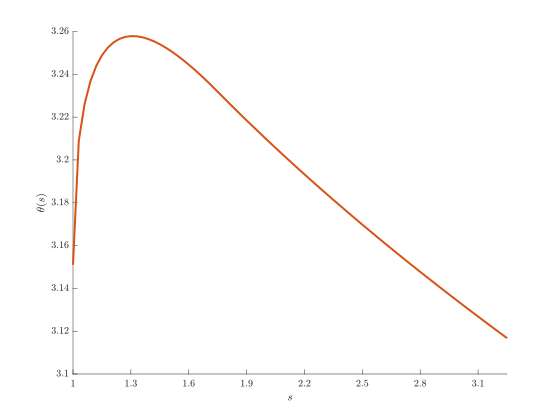}
      \caption{$\theta(s)$ vs $s$ for $N_A=2$, $N_{B}=3$, $N_{E}=1$ and $P=10$ for a particular $H_{B}$ and $H_{E}$.}
      \label{fig:theta_ex}
   \end{figure}

We wish to fix values of $s$, where $s=s(Q)$ for some $Q$, and perform a convex optimisation for $f(Q)$ given this constraint. We then wish to take the maximum value of $f(Q)$ over all values of $s$. Therefore we define $\theta(\cdot)$ as:
\begin{equation}\label{eq:theta}\theta(s)=\max_{Q:s(Q)=s}f(Q).\end{equation}
A plot of $\theta(s)$ can be seen in Figure \ref{fig:theta_ex}. Motivated by the apparent concavity of the simulation results, we aim to prove the concavity regions of these curves.
The simulations and figures presented in this chapter runs the optimisation presented above for fixed values of $s$ using convex optimisation software \emph{CVX: Matlab Software for Disciplined Convex Programming}~\cite{cvx} but the theory holds for an arbitrary convex solver.

Finding the secrecy capacity is now a case of finding the maximum of $\theta(s)$. This is facilitated by the following Theorem,
which gives a concavity result for $\theta$ which is the main result of our paper~\cite{ICC_CJP}.

Let $Q_{i}$ be a matrix achieving the maximum value in Equation \eqref{eq:theta} corresponding to $s_{i}$, that is $f(Q_{i})=\theta(s_{i})$, for $i\in\{1,2\}$. By definition
\begin{align}s_{i}=I_{N_{E}}+ H_{E}Q_{i}H_{E}^{*}
\end{align}
where the $\det$ is no longer required since $N_E = 1$.
Without loss of generality, assume that $s_{1}\geq s_{2}$. Let $s_{t}$ be a convex combination of $s_{1}$ and $s_{2}$
    \begin{align}\label{eq:convex_s}
    s_{t}=ts_{1}+(1-t)s_{2}
    \end{align}
for $t\in [0,1]$.
\begin{thm} \label{thm:main}
For $N_{E}=1$ and any $N_{B}\geq N_{A}$, then
    \begin{align}\label{eq:convex_theta}
    \theta(s_t) \geq t \theta(s_1) + (1-t) \theta(s_2),
    \end{align}
if the matrices $K_B$ and $K_E$ from Equations \eqref{eq:kbdef} and \eqref{eq:kedef} satisfy
    \begin{align}
    &\frac{s_1}{\| K_B^{-1} K_E^2 K_B^{-1} \|_F} - 1 \geq \max\{\lambda_{\max}( H_{B}Q_{1}H_{B}^{*}),\lambda_{\max}( H_{B}Q_{2}H_{B}^{*})\}.
    \end{align}
\end{thm}
\subsection{Overlap with existing results}\label{sec:mimo_result_overlap}
For the antenna configuration $N_{E}=1$, $N_{B}\geq N_A$ required for Theorem \ref{thm:main} to hold there is only one fully understood case. This is the `(2,2,1)' case~\cite{221}, where $N_A=2$, $N_B=2$ and $N_E=1$.
\begin{eg}[2,2,1]
Figure \ref{fig:mimo_221} shows that the theoretical secrecy capacity found in~\cite{221} matches the maximum value of $\theta(s)$.
\begin{figure}[!h]
    \centering
    \includegraphics[width=\textwidth]{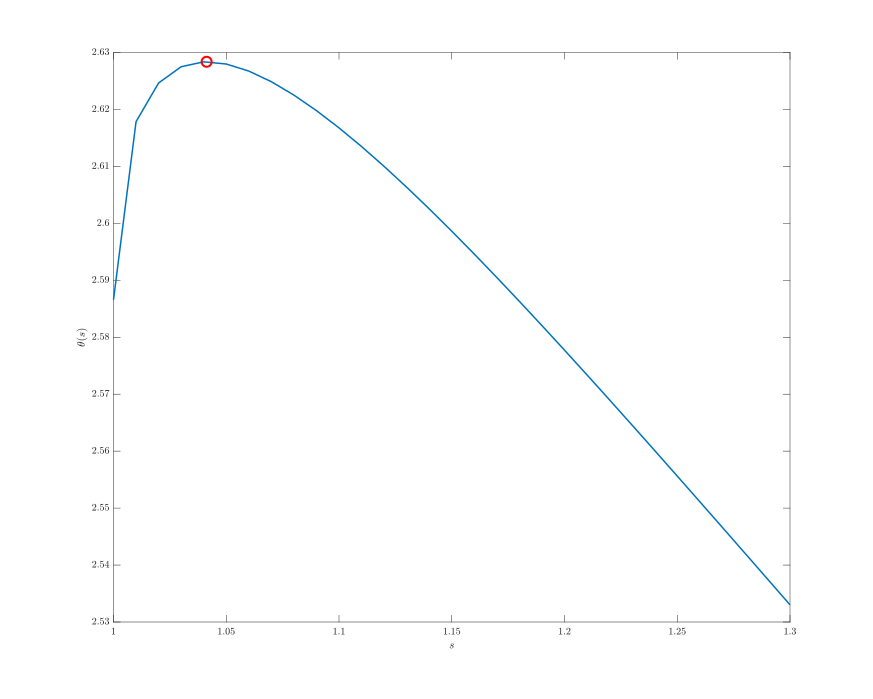}
    \caption{$\theta(s)$ vs $s$ for the (2,2,1) case. The red mark indicates the theoretical secrecy capacity from the paper \cite{221}.}
    \label{fig:mimo_221}
\end{figure}
\end{eg}

\section{Proof of the concave region}\label{sec:proof}
The main argument in the proof of Theorem \ref{thm:main} involves a Taylor expansion of a matrix term which is then bounded at the second order.
The proof can be broken down into three key steps as follows.

\begin{enumerate}
\item Firstly, we consider the function $\theta(s)$, defined in Equation \eqref{eq:theta}, for a convex combination of inputs, $s_t$ (Equation \eqref{eq:convex_s}). Using Lemma \ref{lem:courtade}, which is a second order concavity bound for the $\log\det$, we find a lower bound for $\theta(s_{t})$.
\item We then minimise the difference between the bound from Step 1 with the lower bound required for concavity.
\item Finally, we rewrite these bounds in terms of symmetric matrices which allows us to exploit properties of the Frobenius norm resulting in the conditions stated in Theorem \ref{thm:main}.
\end{enumerate}
\subsection{Step 1}
In this step of the proof, concavity results from~\cite{courtade_etal} are applied to the function $\theta(\cdot)$ defined in Equation \eqref{eq:theta}.
The use of these results allows us to find a tighter lower bound than the usual concavity lower bounds.
\begin{lem}{Courtade et al.~\cite[Lemma 15]{courtade_etal}}\label{lem:courtade}
For positive definite matrices $A$ and $B$ and for any $t\in [0,1]$
\begin{align}\label{courtade}  \log \det(tA+(1-t)B) 
& \geq t\log\det(A)+(1-t)\log\det(B)\nonumber\\
& + \frac{t(1-t)}{2\max\{\lambda_{\max}^{2}(A),\lambda_{\max}^{2}(B)\}}\norm{A-B}_{F}^{2},\end{align}
where $\lambda_{\max}(\cdot)$ denotes the largest eigenvalue and $\norm{\cdot}_{F}$ is the Frobenius norm.
\end{lem}
\noindent For ease of notation, define \begin{equation}\CC_{\max}(A,B)=\frac{\norm{A-B}_{F}^{2}}{2\max\{\lambda_{\max}^{2}(A),\lambda_{\max}^{2}(B)\}}.\end{equation}
Considering the linear combination $Q_{t}=tQ_{1}+(1-t)Q_{2}$, it can be seen that $Q_{t}$ satisfies the constraint $s(Q_{t})=s_{t}$ since $N_{E}=1$ and
\begin{align*}s_{t}=&ts_1+(1-t)s_2\\
=&t(I_{N_{E}}s_{1}+H_{E}Q_{2}H_{E}^{*})
+(1-t)(I_{N_{E}}s_{2}+H_{E}Q_{2}H_{E}^{*})\\
=&I_{N_{E}}+H_{E}(tQ_{1}+(1-t)Q_{2})H_{E}^{*}\\
=&I_{N_{E}}+H_{E}Q_{t}H_{E}^{*}\\
:=&s(Q_{t}).
\end{align*}
Hence
\begin{align}
\theta(s_{t}) \geq f(Q_{t}).
\end{align}
By Lemma \ref{lem:courtade}, taking $A=I_{N_{B}}+H_{B}Q_{1}H_{B}^{*}$ and $B=I_{N_{B}}+H_{B}Q_{2}H_{B}^{*}$ then $f(Q_t)$ is bounded below as follows.
\begin{align}\label{init_lb}
f(Q_{t}) = &\log\det(I_{N_{B}}+H_{B}Q_{t}H_{B}^{*})-\log s_{t}\\
\geq & t\log\det(I_{N_{B}}+H_{B}Q_{1}H_{B}^{*})  + (1-t)\log\det(I_{N_{B}}+H_{B}Q_{2}H_{B}^{*})\nonumber\\ 
&-\log s_{t} + t(1-t)\CC_{\max}A,B).\nonumber
\end{align}
Rewriting the lower bound in Equation \eqref{init_lb} gives
\begin{align}\label{eq:lb_rearrange}
t& (\log\det(I_{N_{B}}+H_{B}Q_{1}H_{B}^{*})-\log s_{1}) \nonumber\\
& + (1-t)(\log\det(I_{N_{B}}+H_{B}Q_{2}H_{B}^{*})-\log s_{2}) + t(1-t)\CC_{\max}A,B) \nonumber\\
&  + t\log s_{1} + (1-t)\log s_{2} -\log(ts_{1}+(1-t)s_{2}) \nonumber \\
=& tf(Q_{1}) + (1-t)f(Q_{2}) + t(1-t)\CC_{\max}A,B) \nonumber \\
& + t\log s_{1} + (1-t)\log s_{2} -\log(ts_{1}+(1-t)s_{2})  \nonumber \\
=& t \theta(s_1) + (1-t) \theta(s_2) + t(1-t)\CC_{\max}A,B) \nonumber \\
&  + t\log s_{1} + (1-t)\log s_{2} -\log(ts_{1}+(1-t)s_{2}), 
\end{align}
since each of the $Q_i$ are optimal by definition.


\subsection{Step 2}


In this step, we aim to minimise the difference between \begin{align*}tf(Q_{1}) + (1-t)f(Q_{2})\end{align*} in Equation \eqref{eq:lb_rearrange} and the upper bound, $\theta(s_{t})$ as defined in Equation \eqref{eq:theta}. To do this, we introduce a constant $\kappa(s_{1},s_{2})$ and show that the following Lemma holds.
\begin{lem}\label{lem:kappa}
For $t\in[0,1]$,
\begin{align}
t\log(s_{1}) + & (1-t)\log(s_{2}) - \log(ts_{1}+(1-t)s_{2}) \nonumber \\
& \geq - t(1-t)\kappa(s_{1},s_{2}),
\label{eg:g_inequality}
\end{align}
for \begin{align}\label{kappa}\kappa(s_{1},s_{2}) = \frac{(s_{1}-s_{2})^{2}}{2s_{1}^{2}}.\end{align}
\end{lem}
\begin{proof}
Define a function $g$ as:
\begin{align}\label{eq:gdef}g(t):=&t\log(s_{1})+(1-t)\log(s_{2})  - \log(ts_{1}+(1-t)s_{2})+ t(1-t)\kappa(s_{1},s_{2})\end{align}
where $\kappa(\cdot,\cdot)$ is a constant. We wish to show that $g(t)\geq0$ for all $t\in [0,1]$.

By construction, $g(0)=g(1)=0$ and therefore $g(t)\geq 0$ in the interval $t\in [0,1]$ is equivalent to $g(t)$ being concave in this interval or when $g''(t)\leq 0$.

The second derivative of $g$ with respect to $t$ is:
$$g''(t)=-2\kappa(s_{1},s_{2})+\frac{(s_{1}-s_{2})^{2}}{s_{t}^{2}}.$$
Since $s_{2}\leq s_{1}$, $g(t)$ is concave for the value of $\kappa(s_{1},s_{2})$ in Equation \eqref{kappa} and thus $g(t)\geq0$ on the interval.
\end{proof}

\subsection{Step 3}
Combining Lemma \ref{lem:kappa} with Equation \eqref{courtade}, we see that Theorem \ref{thm:main} will follow from Equation \eqref{eq:lb_rearrange}
if
\begin{align}\label{eq:seccap_constraints}
\frac{\norm{A-B}_{F}^{2}}{2\max\{\lambda_{\max}^{2}(A),\lambda_{\max}^{2}(B)\}} &\geq \kappa(s_{1},s_{2})\nonumber \\
&\geq \frac{(s_{1}-s_{2})^{2}}{2s_{1}^{2}}
\end{align}
where, as before,
\begin{equation}\label{matrixA}A := I_{N_{B}} + H_{B}Q_{1}H_{B}^{*}\end{equation}
and
\begin{equation}\label{matrixB}B := I_{N_{B}} + H_{B}Q_{2}H_{B}^{*}.\end{equation}
Writing $\overline{Q}:=Q_{1}-Q_{2}$ for simplicity, 
the Frobenius norm on the left of Equation \eqref{eq:seccap_constraints} can be rewritten as
\begin{align}
\norm{A-B}_{F}^{2}&= \tr(H_{B}\overline{Q}H_{B}^{*}H_{B}\overline{Q}H_{B}^{*})\nonumber\\
&= \tr(\overline{Q}K_{B}^{2}\overline{Q}K_{B}^{2})\nonumber\\
&= \tr((K_{B}\overline{Q}K_{B})(K_{B}\overline{Q}K_{B}))\nonumber\\
&=\tr(RR)=\tr(RR^{*})\nonumber\\&=\norm{R}^{2}_{F}
\end{align}
where $R$ is the symmetric matrix
\begin{align}
R& :=K_{B}\overline{Q}K_{B}.\end{align}
In order to retrieve the value of $\overline{Q}$ from $R$ requires that $H_{B}^{*}H_{B}$ is invertible. This implies that $N_{B}\geq N_A$.

Similarly, considering the numerator of the right hand side of Equation \eqref{eq:seccap_constraints} gives:
\begin{align}
(s_{1}-s_{2})^{2}&=(H_{E}\overline{Q}H_{E}^{*})^{2}\nonumber\\
&= \tr(\overline{Q}K_{E}^{2}\overline{Q}K_{E}^{2})\nonumber\\
&= \tr((K_{E}\overline{Q}K_{E})(K_{E}\overline{Q}K_{E}))\nonumber\\
&= \tr(RTRT) \nonumber \\
&\leq\norm{RT}^{2}_{F} \label{eq:tojustify1} \\
&\leq \norm{R}_{F}^{2}\norm{T}_{F}^{2}. \label{eq:tojustify2}
\end{align}
where $T$ is the symmetric matrix
\begin{align}
T& :=K_{B}^{-1}K_{E}^{2}K_{B}^{-1}.\end{align}
Here, Equation \eqref{eq:tojustify1} follows
by Cauchy-Schwarz, for any matrix $C$,
$$ \tr(C^2)\leq \tr(C^{*}C)=\norm{C}^{2}_{F},$$
and Equation \eqref{eq:tojustify2} follows by the submultiplicative property of the Frobenius norm.
Since both $R$ and $T$ are symmetric, the following holds:
\begin{align}
\tr(RTRT)&\leq\norm{RT}^{2}_{F}.
\end{align}
\begin{align}
(s_{1}-s_{2})^{2}&\leq \norm{R}_{F}^{2}\norm{T}_{F}^{2}.
\end{align}
Therefore the inequality in Equation \eqref{eq:seccap_constraints} is satisfied when
\begin{align}\label{eq:inequality_frobnorm}
\frac{\norm{R}^{2}_{F}}{2\max\{\lambda_{\max}^{2}(A),\lambda^{2}_{\max}(B)\}}\geq \frac{\norm{R}^{2}_{F}\norm{T}^{2}_{F}}{2s_{1}^{2}}.
\end{align}
Since each of $\lambda_{\max}^{2}(\cdot)$, $\norm{T}_{F}^{2}$ and $s_{1}^{2}$ is positive, it is possible to present the conditions for satisfying Equation \eqref{eq:inequality_frobnorm} as follows:
\begin{align}\label{eq:sq_ineq}
s_{1}\geq \max\{\lambda_{\max}(A),\lambda_{\max}(B)\}\norm{T}_{F},
\end{align}
and the proof of Theorem \ref{thm:main} is complete.
%
%
%
%
%

\section{Outside the concave region}\label{sec:notconcave}
The function $\theta(\cdot)$ cannot be concave indefinitely, since the secrecy capacity must be non negative by definition. A negative secrecy capacity would be a worse regime than sending nothing, and thus a rate of 0 would be preferable.
We wish to show that the function does not have another maximum, and therefore the maximum found in Equation \ref{eq:theta} is the true secrecy capacity. If we can show that there exists a cutoff, $a$, such that $\theta(\cdot)$ is concave on $[0,a)$, is convex for $(a,\infty)$ and tends to 0 then this is sufficient.

In the proof of Theorem \ref{thm:main}, Lemma \ref{lem:courtade} was used to find a lower bound for $\log\det(\cdot)$. In this section, we prove a converse of Lemma \ref{lem:courtade} and then apply this to $\theta(\cdot)$.

Firstly, we define an $M(x,y)$-strongly concave function, and then apply this definition to $\log\det(\cdot)$ to find an upper bound on the log determinant of a convex combination of arguments (analogous to the lower bound in \cite[Lemma 15]{courtade_etal}). This is then applied in a similar manner to the proof of Theorem \ref{thm:main} to give a result about $\theta(\cdot)$ outside of the concave region.

\begin{defn}\label{def:stronglyconcave}
A twice differentiable function $f : \dom f \to \mathbb{R}$ is $M(x,y)$-strongly concave between $x,y\in \dom f$ if $\nabla^2 f(tx+(1-t)y) \leq M(x,y)I$ for all $t \in [0,1]$.
\end{defn}

\begin{lem}\label{lem:Mstrongconcave}
For all $t \in [0,1]$, an $M(x,y)$-strongly concave function $f$ satisfies
\begin{align}
    tf(x) + (1-t)f(y)& \leq f(tx+(1-t)y) + t(1-t)\frac{M(x,y)}{2}| x-y |^2.
\end{align}
\end{lem}

The proof of this lemma is largely the same as the proof of [Lemma 30]\cite{courtade_etal} but tackles the problem from the other side (that is, to give an upper bound rather than their lower bound).

\begin{proof}
The Taylor series expansion of $f$ for any two points $x,y\in \text{dom} f$ yields
\begin{align}
    f(x) = & f(y) + \langle\nabla f(y),y-x\rangle \nonumber\\
           & + \frac{1}{2}\langle y-x,\nabla^2 f(t_0 a + (1-t_0)b)(y-x)\rangle \label{eq:taylorexp}\\
      \leq & f(y) + \langle\nabla f(y), y-x\rangle + \frac{M(x,y)}{2}| y-x | ^2, \label{eq:inequal_mconcave}
\end{align}
where Equation \eqref{eq:taylorexp} holds for some $t_0\in [0,1]$ and Equation \eqref{eq:inequal_mconcave} follows from Definition \ref{def:stronglyconcave}.
Let $w = tx+(1-t)y$, for $t\in [0,1]$. Then applying the above inequality to $f(x)$ and $f(y)$ gives
\begin{align}
    f(x) & \leq f(w) + \langle\nabla f(w),w-x\rangle + \frac{M(w,x)}{2}| w-x | ^2\label{eq:concaveproof1}\\
    f(y) & \leq f(w) + \langle\nabla f(w),w-y\rangle + \frac{M(w,y)}{2}| w-y | ^2.\label{eq:concaveproof2}
\end{align}
Summing $t$\eqref{eq:concaveproof1} + $(1-t)$\eqref{eq:concaveproof2} yields
\begin{align}\label{eq:concaveproof3}
    tf(x) + (1-t)f(y) & \leq f(w) + \frac{t(1-t)^2M(x,w) + t^2(1-t)M(y,w)}{2}| y-x | ^2.
\end{align}

By definition of $w$, $M(x,w)\leq M(x,y)$ and $M(y,w)\leq M(x,y)$ and therefore Equation \eqref{eq:concaveproof3} may be bounded above by
\begin{align}
    f(tx+(1-t)y) + t(1-t)\frac{M(x,y)}{2}| y-x | ^2
\end{align}
which proves the lemma.
\end{proof}

We now give an upper bound for $\log\det(\cdot)$ for convex combinations, this is analagous to the lower bound of Lemma \ref{lem:courtade}.

\begin{lem}\label{lem:logdet_strongconvex}
    For positive definite matrices $A$, $B$ and $t\in[0,1]$,
    \begin{align}
        \log\det (tA + (1-t)B) \leq & t\log\det (A) + (1-t) \log\det(B)\nonumber\\
        & + \frac{t(1-t)}{2\min \{\lambda^2_{\min}(A),\lambda^2_{\min}(B)\}}\norm{A-B}^{2}_{F}.       
    \end{align}
    For ease of notation, we denote
    \begin{align}
        \CC_{\min}(A,B) = \frac{1}{2\min \{\lambda^2_{\min}(A),\lambda^2_{\min}(B)\}}.
    \end{align}
\end{lem}

Again, the proof closely follows that given in \cite{courtade_etal} for their equivalent Lemma, but uses the concavity of $\log\det(\cdot)$ rather than the convexity of $-\log\det(\cdot)$.

\begin{proof}
    Since $f(\cdot) = \log\det (\cdot)$ is strictly concave and twice differentiable for positive semidefinite matrices, we may apply Lemma \ref{lem:Mstrongconcave} to $f$.
    Therefore
    \begin{align}
        \log\det (tA + (1-t)B) \leq t\log\det (A)& + (1-t) \log\det(B)\nonumber\\
        & + t(1-t)\frac{M(A,B)}{2}\norm{A-B}^{2}_{F}. \label{eq:logdet_Mconcave}
    \end{align}
    
    Since $\nabla^2 f(C) = C^{-1}\otimes C^{-1}$, where $\otimes$ denotes the Kronecker product.
    The maximum eigenvalue of this product is given by $1/\lambda_{\min(C)}$
    (since eigenvalues of $X\otimes Y$ are the products of eigenvalues of $X$ and eigenvalues of $Y$.).
    By the definition of $M(A,B)$ we have the following upper bounds
    \begin{align}
        M(A,B) & \leq \max_{t\in [0,1]} \frac{1}{\lambda^2_{\min} \left(tA + (1-t)B\right)}\label{eq:M_upperbound}\\
        & \leq \frac{1}{\min \{\lambda^2_{\min}(A),\lambda^2_{\min}(B)\}},\label{eq:M_evalupperbound}
    \end{align}
    \vspace{3pt}
    where Equation \eqref{eq:M_evalupperbound} follows by the concavity of the minimum eigenvalue.
    Combining Equations \eqref{eq:logdet_Mconcave} and \eqref{eq:M_evalupperbound} gives
    \begin{align}
        \log\det (tA + (1-t)B) \leq & t\log\det (A) + (1-t) \log\det(B)\\
        & + t(1-t)\frac{1}{\min \{\lambda^2_{\min}(A),\lambda^2_{\min}(B)\}}\norm{A-B}^{2}_{F}\nonumber
    \end{align}
    as desired.
\end{proof}

Using definitions and properties, we give a result describing the behaviour of $\theta(\cdot)$ outside the concave region.

\begin{thm}\label{thm:thetaconvexity}
Given $s_t$, let $Q_t$ be the corresponding optimal covariance matrix.
Then
\begin{align}
    \theta(s_{t}) \leq t\theta(s(Q_{1})) + (1-t)\theta(s(Q_{2}))
\end{align}
for any positive semidefinite matrices $Q_{1}$, $Q_{2}$ such that $Q_{t} = tQ_{1} + (1-t)Q_{2}$ if the matrices $K_B$ and $K_E$ from Equations \eqref{eq:kbdef} and \eqref{eq:kedef} satisfy
    \begin{align} \label{eq:cond}
    &\frac{s(Q_{1})}{\| K_B^{-1} K_E^2 K_B^{-1} \|_F} - 1 \leq \min\{\lambda_{\min}( H_{B}Q_{1}H_{B}^{*}),\lambda_{\min}( H_{B}Q_{2}H_{B}^{*})\}.
    \end{align}
\end{thm}

\begin{figure}[!h]
    \centering
    \includegraphics[width=\textwidth]{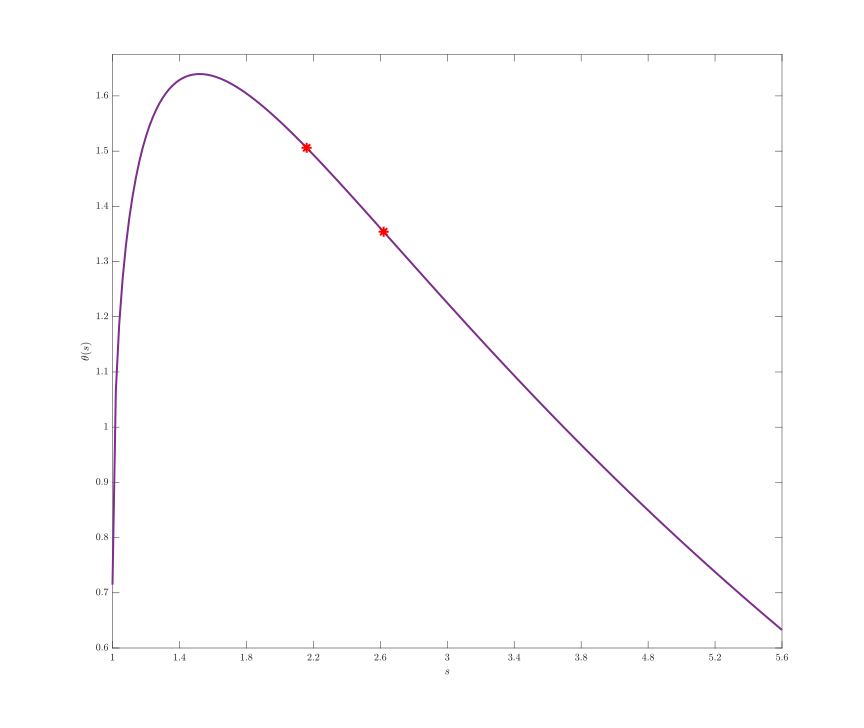}
    \caption{$\theta(s)$ vs $s$ for a particular channel, showing the cut off points for the inequalities in Theorems \ref{thm:main} and \ref{thm:thetaconvexity}.}
    \label{fig:mimo_scutoff}
\end{figure}

Simulation results showing the cutoff points for the convex and concave regions can be seen as red markers in Figure \ref{fig:mimo_scutoff}. It can be seen that there is a gap between these two, and this is expected since in the proofs some conservative bounds are applied however, this is only a small region which can easily be searched across.

Analogously to the proof of the concave region, the main steps of this proof will be roughly the same.
\begin{enumerate}
    \item Finding an upper bound for $f(Q_t)$ using Theorem \ref{lem:logdet_strongconvex}.
    \item Minimising the difference between the bound found in the first step with the desired convexity bound.
    \item Rewriting these bounds in terms of symmetric matrices and applying properties of the Frobenius norm.
\end{enumerate}

\subsection{Step 1}
Let $Q_i$ denote the matrix which achieves the maximum value of $\theta$ for $s_i$. Recall the definition of $s(Q_{i}) = I_{N_E} + H_{E}Q_{i}H_{E}^{*}$, and for optimal $Q_{i}$, we have that $s(Q_{i}) = s_{i}$.
Choose $s_t$ and the corresponding $Q_{t}$.
For some $t \in [0,1]$, write
\begin{align}\label{eq:st_convex}
    s_t = ts(\olQ_{1}) + (1-t)s(\olQ_{2})
\end{align}
($s(\olQ_{1})\geq s(\olQ_{2})$) where $Q_{t} = t\olQ_{1} + (1-t) \olQ_{2}$. Here we use the notation $\olQ_{i}$ to distinguish this matrix from the optimal matrix $Q_{i}$.

By applying Lemma \ref{lem:logdet_strongconvex}, and using the optimality of $Q_{t}$ we may bound $\theta(s_{t})$ as follows
\begin{align}
    \theta(s_{t}) = & f(Q_t) = \log\det (I+H_{B}Q_{t}H_{B}^{*}) - \log s_{t} \nonumber\\
        \leq & t\log \det (A) + (1-t)\log\det(B) + t(1-t)\CC_{\min}(A,B) - \log s_{t} \label{eq:f_upperbound}
\end{align}
for $A= I + H_{B}\olQ_{1}H_{B}^{*}$ and $B = I + H_{B}\olQ_{2}H_{B}^{*}$.
Equivalently, the upper bound of Equation \eqref{eq:f_upperbound} may be bounded by
\begin{align}
    tf(\olQ_1)& + (1-t)f(\olQ_2) + t(1-t)\CC_{\min}(A,B)\nonumber\\
    & +  t\log(s_{1}) + (1-t)\log (s_{2}) - \log (s_{t})\label{eq:ftheta1}\\
     \leq t\theta(s_{1})& + (1-t)\theta(s_{2}) + t(1-t)\CC_{\min}(A,B) \nonumber\\
     t\log(s_{1}) & + (1-t)\log (s_{2}) - \log (s_{t})\label{eq:ftheta2}
\end{align}
where Equation \eqref{eq:ftheta2} follows by the definition of $Q_{1,2}$ and the fact that $\theta(\cdot)$ is a maximum.

\subsection{Step 2}
We require the following Lemma.
\begin{lem}\label{lem:g_convex}
    For all $t\in [0,1]$,
    \begin{align}
        t\log(s_1) +& (1-t)\log(s_2) - \log (ts_{1} + (1-t)s_{2}) \nonumber\\
        & \leq -t(1-t)\kappa(s_{1},s_{2}), \label{eq:logs_ubound}
    \end{align}
    for
    \begin{align}
        \kappa(s_1,s_2) = \frac{(s_{1} - s_{2})^{2}}{2s_{1}^{2}}.
    \end{align}
\end{lem}

\begin{proof}
    Following the proof of Lemma \ref{lem:kappa}, this is a matter of showing that the equivalent function $g$ is convex in the interval for this value of $\kappa$.
\end{proof}

\subsection{Step 3}
The desired convexity constraints follows by combining Lemma \ref{lem:g_convex} and Lemma \ref{lem:logdet_strongconvex}. The desired bound is held if
\begin{align}
    \frac{t(1-t)}{2\min \{ \lambda_{\min}^{2}(A),\lambda_{\min}^{2}(B)\}} &\leq \kappa(s_{1},s_{2})\\
    & \leq \frac{(s_{1}-s_{2})^{2}}{2s_{1}^{2}}.
\end{align}
And so the result follows by the definitions of $R$ and $T$ given in the proof of Theorem \ref{thm:main}.

\section{Discussion}\label{sec:discussion_mimo}

Although the expression for the secrecy capacity is known for the Gaussian wiretap channel, it is not generally known how to solve the optimisation problem for the covariance matrix, $Q$. The method presented in this chapter gives an efficient way to search for the secrecy capacity of a MIMO system and a corresponding covariance matrix for the transmission.
The use of existing convex optimisation schemes makes the problem presented in Equation \eqref{eq:ch_MIMO_seccap} manageable. We show that it is possible to efficiently search numerically for the maximum using linear combinations of variables.

For a fixed channel, the norm $\norm{T}_{F}$ is simple to compute. To find the secrecy capacity, it is a case of picking a value of $s_1$ and $s_2$ and checking the constraint in Equation \eqref{eq:sq_ineq}. If the criteria is satisfied, then these are in the concave region. It is therefore sufficient to use a standard concave optimisation technique. If Equation \eqref{eq:sq_ineq} is not satisfied, then an algorithm may be implemented to choose a different value until we are in the concave region.

The transmission scheme corresponding to this covariance matrix will be information theoretically secure since the user is guaranteed to be transmitting at or below the secrecy capacity.

This scheme is specific to the case with $N_{E}=1$ and $N_{B}\geq N_A$. This is due to the requirements which arise in the derivation of the proof. Despite these restrictions, this work covers a family of MIMO systems which are not fully understood at the time of writing. For the situation with multiple antennas at the eavesdropper, the current state of the art is the algorithmic approach outlined by~\cite{loyka2015algorithm}. When the number of antennas at Eve is greater than 1, the problem of the secrecy capacity cannot be written in the equivalent convex format as outlined in this chapter and the problem becomes far more difficult. In the Gaussian setup, multiple single antenna eavesdroppers behave in the same way as a multiple antenna eavesdropper. It is unclear whether this helps in this particular scenario, but is an avenue for future investigation.

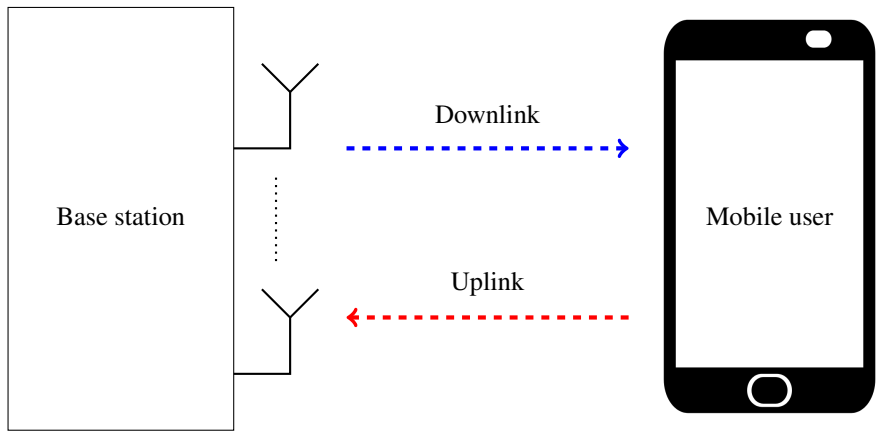
\begin{figure}
    \centering
    \begin{tikzpicture}[scale=1.5]
    \draw (0,0) rectangle (2,3.75);
    \node at (1,1.9) {Base station};
    \draw [thick] (2,0.5) -- (2.5,0.5) -- (2.5,1) -- (2.25,1.25);
    \draw [thick] (2.5,1) -- (2.75,1.25);
    \draw[thick, dotted] (2.375,1.5) -- (2.375,2.25);
    \draw [thick] (2,2.5) -- (2.5,2.5) -- (2.5,3) -- (2.25,3.25);
    \draw [thick] (2.5,3) -- (2.75,3.25);
    \node[inner sep=0pt] (MU) at (6.75,1.9)
    {\includegraphics[width=3cm]{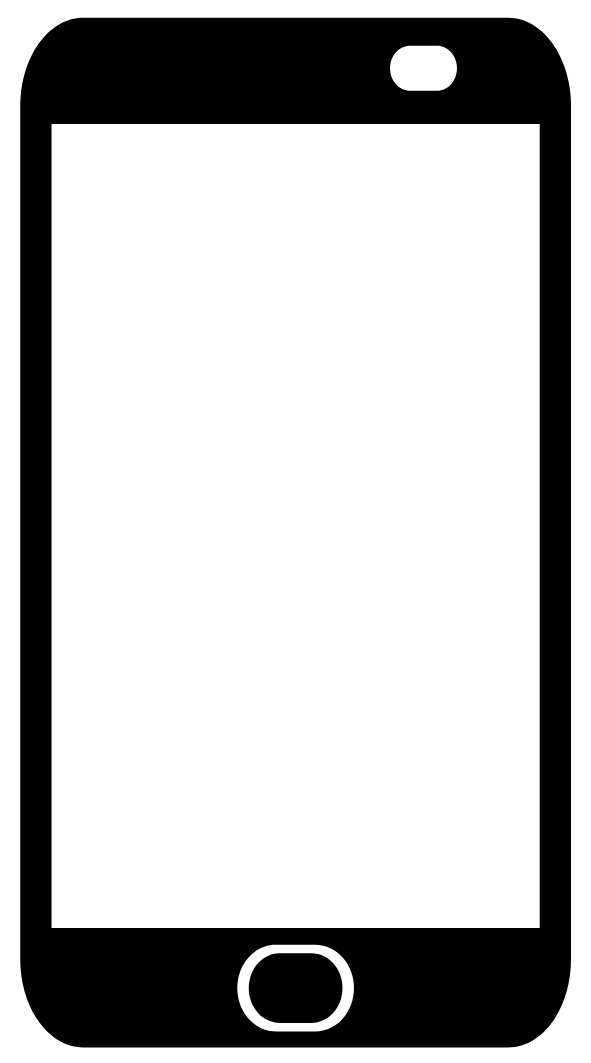}};
    \node at (6.75,1.9) {Mobile user};

    \draw [ultra thick, red,dashed, ->] (5.5,1) -- (3,1);
    \draw [ultra thick, blue,dashed, ->] (3,2.5) -- (5.5,2.5);
    \node (uplink) at (4.25,1.3)
    {Uplink};
    \node (downlink) at (4.25,2.8)
    {Downlink};
    \end{tikzpicture}

    \caption[A massive MIMO basestation and mobile user]{Massive MIMO basestations will have a far greater number of antennas than the mobile users}
    \label{fig:basestation_uplink}
\end{figure}

 In order to achieve the desired capacity gains for 5G, massive MIMO systems are a key technology~\cite{5Gandrews}.
 This means that modern and future systems using massive MIMO will have a high number of antennas at the base station. Therefore the $N_B \geq N_A$ constraint in Theorem \ref{thm:main} would imply that these results are limited to the uplink for a massive MIMO system, as in Figure \ref{fig:basestation_uplink} since mobile users will have far fewer antennas. In future work, it would be interesting to generalise to the downlink of such channels.


It is important to note that this work assumes a static environment. Since the work considers the Gaussian wiretap channel with full channel state information (CSI), there is an inherent assumption that the channel statistics are fixed. Thus these results hold within the coherence time of the channel therefore the channel is fairly static, they are valid for a longer period of time.
If we no longer assume a static channel, and instead suppose that the channel matrices are unknown, or fading, then the dimensions of the problem increase dramatically. For different types of fading, Equation \eqref{eq:ch_MIMO_seccap} is no longer the agreed formula for the secrecy capacity, and there are far more degrees of freedom in the problem.

A practical limitation of any capacity result stemming from Shannon's work is the asymptotic nature of the results. While it is important to understand the fundamental measures of systems, there is evidence that the capacity of a system may be significantly lower for finite blocklength as shown in~\cite{polyanskiy_blocklength}. This means that the secrecy capacity could be an overestimate, particularly for low power devices with short blocklength such as internet of things devices.

Theorem \ref{thm:thetaconvexity} is a weaker statement than that in Theorem \ref{thm:main}. This is because in finding the upper bound, and thus the concavity of $\theta(\cdot)$, firstly $s_1$ and $s_2$ are picked and then a convex combination $s_t = t s_{1} + (1-t) s_{2}$ is taken. Since $\theta(\cdot)$ is a maximum of $f(Q)$ taken over all $Q$, we may upper bound our statement by $\theta(s_{t})$. In the proof of Theorem \ref{thm:thetaconvexity}, firstly $s_t$ is picked. From here, it is not immediate that a value of $s_1$ and $s_2$ exist under the given constraints, and so the Theorem statement is looser.

\section{Acknowledgments}\label{sec:acknowledgements}

This work was supported by the Engineering and Physical
Sciences Research Council [grant number EP/I028153/1]; GCHQ;
and the University of Bristol.

\bibliographystyle{IEEEtran}
\bibliography{bibliography.bib}

\end{document}